**Electron-Hole Asymmetry of Spin Injection and Transport in Single-Layer Graphene**

Wei Han, W. H. Wang,[†] K. Pi, K. M. McCreary, W. Bao, Yan Li, F. Miao, C. N. Lau, and R. K. Kawakami[‡]

Department of Physics and Astronomy, University of California, Riverside, CA 92521

[†]Present address: Institute of Atomic and Molecular Sciences, Academia Sinica, Taipei 106, Taiwan.
[‡]e-mail: roland.kawakami@ucr.edu

**Abstract:**

Spin-dependent properties of single-layer graphene (SLG) have been studied by non-local spin valve measurements at room temperature. Gate voltage dependence shows that the non-local magnetoresistance (MR) is proportional to the conductivity of the SLG, which is the predicted behavior for transparent ferromagnetic/nonmagnetic contacts. While the electron and hole bands in SLG are symmetric, gate voltage and bias dependence of the non-local MR reveal an electron-hole asymmetry in which the non-local MR is roughly independent of bias for electrons, but varies significantly with bias for holes.

Graphene is an attractive material for spintronics due to its tunable carrier concentration [1-3], weak spin-orbit coupling, predictions of novel spin-dependent behavior [4, 5], and the recent experimental observations of spin transport [6-12]. A special property of single-layer graphene (SLG) is that the band structure of the electrons and holes are ideally symmetric (similar to carbon nanotubes [13]), so their spin-dependent properties are expected to match. This differs from conventional semiconductors such as GaAs and Si, whose electron and hole bands are highly asymmetric (e.g. different atomic orbital states, different spin-orbit coupling, different effective masses), which leads to very different spin-dependent properties. Thus, the observation of electron-hole asymmetry of a spin-dependent property in SLG would create a unique opportunity to investigate the relationship between carrier charge and spin, separated from the typical effects of band asymmetries found in conventional semiconductors.

In this Letter, we report the observation of electron-hole asymmetry for spin injection and transport in SLG at room temperature, as determined by non-local magnetoresistance (MR) measurements on SLG spin vales with transparent Co contacts [14, 15]. A systematic investigation of the gate voltage dependence and bias dependence of the non-local MR signal shows that when the carriers in the SLG are electrons, the non-local MR is roughly constant as a function of dc current bias, which is consistent with the standard one dimensional (1D) drift-diffusion model of spin injection and transport [15-19]. When the carriers in the SLG are holes, however, the non-local MR is strongly reduced in the negative bias regime (i.e. spin extraction [20]). This differing behavior between the electrons and the holes is a clear demonstration of spin-dependent electron-hole asymmetry, which is most likely due to an interfacial effect at the Co/SLG contact. Understanding the origin of this asymmetry will be crucial for the development of bipolar spin transport devices utilizing both electrons and holes.

The devices consist of exfoliated SLG sheets [21, 22] and Co electrodes fabricated by electron-beam lithography using PMMA/MMA bilayer resist (Figure 1a). The $SiO_2$/Si substrate (300 nm layer thickness of $SiO_2$) is used as a gate. Because the non-local spin signal should be enhanced by decreasing the contact area [19], we utilize angle evaporation to deposit a 2 nm MgO masking layer prior to the deposition of an 80 nm Co layer (Figure 1a detail). This reduces the width of the contact area to ~50 nm. Prior to lift-off, the device is capped with 5 nm $Al_2O_3$ to protect the Co from further oxidation. For the two representative samples (A and B), the widths of the electrodes are 225 nm, 210 nm, 175 nm, and 225 nm for sample A and 350 nm, 160 nm, 210 nm, and 180 nm for sample B. The spacings between electrodes for sample A are $L_{12}$ = 1.0 μm, $L_{23}$ = 1.0 μm, and $L_{34}$ = 2.0 μm and for sample B are $L_{12}$ = 1.6 μm, $L_{23}$ = 1.0 μm, and $L_{34}$= 1.1 μm. The widths of the SLG are ~2 μm for both samples. Raman spectroscopy is used to verify the thickness of the graphene [23]. Figure 1b shows typical spectra from SLG measured on our devices and from bulk graphite for reference. Figure 1c shows a scanning electron microscope image of a completed device, in which the darker region corresponds to the SLG.

The electrical and non-local magnetoresistance (MR) characteristics are measured in vacuum at room temperature. Figures 2a and 2b show the resistivity of the SLG as a function of gate voltage for samples A and B. Both samples exhibit a peak in resistivity which define the Dirac point, with $V_{Dirac}$ = -34 V for sample A and $V_{Dirac}$ = -32 V for sample B. Sample A has a mobility of 900-1700 cm$^2$/Vs, while sample B has a mobility of 800-1300 cm$^2$/Vs. The *I-V* curves measured across electrodes E1 and E2 at different gate voltages indicate transparent contacts between the Co and SLG (Figures 2c and 2d).

Spin injection and transport are investigated using standard lock-in techniques. A current source applies a dc bias ($I_{dc}$) and ac excitation ($I_{ac}$ = 30 μA) across electrodes E1 and E2 (Figure 1a) to generate spin polarization in the SLG beneath electrode E2 by spin injection or extraction. This spin-polarization propagates to E3 via spin diffusion and generates a non-local voltage across electrodes E3 and E4 ($V = V_{dc} + V_{ac}$) due to the spin-sensitive nature of the ferromagnetic electrodes [14-19]. To separate the spin signal from a constant background level, $R_{NL}$ ($\equiv V_{ac}/I_{ac}$) is measured as the magnetic field is swept up and swept down (Figures 2e and 2f) to generate parallel and antiparallel alignments of the central electrodes (E2 and E3). The non-local MR is defined as $\Delta R_{NL} = R_{NL}^P - R_{NL}^{AP}$, where $R_{NL}^P$ ($R_{NL}^{AP}$) is the non-local resistance for the parallel (antiparallel) state. Figure 2e shows representative non-local MR scans on sample A measured at zero bias. Comparing the scans, we see that $\Delta R_{NL}$ is smallest near the Dirac point ($V_g$ = -30 V) and larger for electron doping ($V_g$ = 0 V) and hole doping ($V_g$ = -70 V). The non-local MR of sample B shows similar behavior, with $\Delta R_{NL}$ smallest when $V_g$ is close to the Dirac point, and higher for larger carrier densities (Figure 2f).

Figures 3a and 3b show the detailed gate-dependence of $\Delta R_{NL}$ at zero bias on samples A and B (circles). $\Delta R_{NL}$ has a minimum near the Dirac point and has increasing values for increasing electron density ($V_g > V_{Dirac}$) as well as for increasing hole density ($V_g < V_{Dirac}$). This behavior can be understood in terms of the 1D drift-diffusion model, which predicts that $\Delta R_{NL}$ should be proportional to the conductivity of the nonmagnetic material, $\sigma_N$, (SLG in our case) for transparent ferromagnetic/nonmagnetic contacts (e.g. equation 4 in ref. [18], equation 1 in ref. [15] with $M \gg 1$). The solid lines in Figures 3a and 3b show the conductivity as a function of gate voltage. The good agreement indicates that we have realized the $\Delta R_{NL} \sim \sigma_N$ dependence for transparent contacts. This illustrates a powerful aspect of graphene as a material to examine spin-polarized transport, where the ability to tune the conductivity provides a novel approach to investigate theoretical predictions.

To gain insight into the characteristics of spin injection and transport in SLG, we systematically investigate the gate dependence and bias dependence of $\Delta R_{NL}$. Figures 3c and 3d show the gate-dependence of $\Delta R_{NL}$ for samples A and B for $I_{dc} = +300$ µA (squares), 0 µA (circles), and -300 µA (triangles). The polarity of $I_{dc}$ is defined in Figure 1a. For positive bias, the gate-dependence of $\Delta R_{NL}$ follows the zero bias data. On the other hand, when the bias is negative and the carriers are holes (triangles, $V_g < V_{Dirac}$), a strong reduction of $\Delta R_{NL}$ is observed in both samples. In this case, the holes in the SLG are driven toward electrode E2 and become spin-polarized due to spin-dependent reflection from the ferromagnetic interface (i.e. spin extraction [20]). A very interesting aspect is that the reduction of $\Delta R_{NL}$ is observed for spin extraction of holes, but not for the spin extraction of electrons.

Figure 4a shows the bias dependence of $\Delta R_{NL}$ on sample A for $V_g = 0$ V (electrons, solid squares) and for $V_g = -70$ V (holes, open squares). For electrons, there is only a slight variation in $\Delta R_{NL}$ as a function of $I_{dc}$. For holes at positive bias, the behavior of $\Delta R_{NL}$ is similar to the electron case. For holes at negative bias, however, there is a significantly stronger variation of $\Delta R_{NL}$ as a function of dc current bias, with decreasing $\Delta R_{NL}$ at larger negative biases. Figure 4c shows the bias dependence of $\Delta R_{NL}$ on sample B for $V_g$ = 10 V (electrons, solid squares) and for $V_g$ = -60 V (holes, open squares). Similar to sample A, for electrons the value of $\Delta R_{NL}$ is roughly constant as a function of dc bias current. For holes under negative bias, there is a very strong change of $\Delta R_{NL}$ with dc current bias, nearly approaching zero at $I_{dc}$ = -300 µA. The images in Figure 4b and 4d show the dependence of $\Delta R_{NL}$ as a function of both gate voltage and dc current bias for samples A and B, respectively. The two main trends, namely the roughly constant $\Delta R_{NL}$ vs. $I_{dc}$ for electrons and the reduced $\Delta R_{NL}$ for hole spin extraction, can be clearly seen in the two images.

The roughly constant $\Delta R_{NL}$ vs. $I_{dc}$ can be understood in terms of the 1D drift-diffusion model [15-19], which predicts that the non-local voltage $\Delta V = \Delta V^P - \Delta V^{AP}$ is proportional to the injection current $I$. For the ac lock-in measurement, this behavior will lead to a constant $\Delta R_{NL}$ vs. $I_{dc}$ because the lock-in measures the slope of the $\Delta V$ vs. $I$ curve. The reduction of $\Delta R_{NL}$ for hole spin extraction represents a deviation from the standard behavior. Similar deviations from the standard behavior have been observed for spin extraction in Fe/n-GaAs [24], CoFe/Al$_2$O$_3$/Al [25], and very recently in Co/Al$_2$O$_3$/graphene [26]. In these studies, tunnel barriers between the ferromagnet and non-magnetic materials play a prominent role in explaining the unusual behavior [20, 25, 27]. In our devices, the contact resistances are less than 300 Ω and have linear *I-V* characteristics, so the behavior is not related to interfacial barriers and must originate from a different physical mechanism. We believe an interfacial effect at the Co/SLG contact such as wavefunction hybridization or local doping could be important [28-31]. With a strong Co-SLG hybridization, it is possible for the spin-dependent density of states of the Co to break the electron-hole symmetry of the SLG [30]. Apart from band structure effects, local doping has been shown to generate electron-hole asymmetry of the conductance [28, 29, 31], but its influence on the spin-dependent properties is currently unclear. Further theoretical and experimental studies will be needed to understand the origin of the electron-hole asymmetry of the spin signal.

In summary, we have measured non-local MR on SLG spin valves as a function of gate voltage and dc current bias. The gate dependence of the non-local MR at zero bias is found to scale with the SLG conductivity, consistent with the predicted behavior for transparent contacts. For electrons, the non-local MR is roughly independent of bias, but for holes under negative bias the non-local MR is strongly reduced. Understanding the origin of this effect should be important for further theoretical developments in spintronics.

We acknowledge the support of ONR (N00014-05-1-0568), NSF (CAREER DMR-0450037), NSF (CAREER DMR-0748910), NSF (MRSEC DMR-0820414), and CNID (ONR/DMEA-H94003-07-2-0703).

FIGURE CAPTIONS:

Figure 1: (a) Schematic diagram of the single layer graphene (SLG) spin valve. E1, E2, E3, are E4 are four cobalt electrodes. The Si substrate acts as a back gate. Detail: A MgO layer deposited by angle evaporation to reduce the width of the contact area to ~50 nm. (b) Raman spectroscopy of SLG and bulk graphite. (c) SEM image of a completed device. The darker region corresponds to the SLG.

Figure 2: Electrical characteristics and non-local magnetoresistance (MR) scans of samples A and sample B. (a,b) SLG resistivity vs. gate voltage of sample A and sample B. (c,d) *I-V* curves between electrodes E1 and E2 of sample A and sample B. (e) Non-local MR scans of sample A at three different gate voltages ($V_g$ = 0 V, -30 V, and -70 V), as the magnetic field is swept up (black curve) and swept down (red curve). A constant background is subtracted and the curves are offset for clarity. (f) Non-local MR scans of sample B at $V_g$ = 10 V, -30 V, and -60 V. A constant background is subtracted and the curves are offset for clarity.

Figure 3: (a) Non-local MR at zero bias (circles) and conductivity (solid line) vs. gate voltage for sample A. (b) Non-local MR at zero bias (circles) and conductivity (solid line) vs. gate voltage for sample B. (c) The dependence of non-local MR on the gate voltage for sample A at bias current 300 µA (squares), 0 µA (circles), -300 µA (triangles). (d) The dependence of non-local MR on the gate voltage for sample B at bias current 300 µA (squares), 0 µA (circles), -300 µA (triangles).

Figure 4: (a) Non-local MR as a function of dc bias current for sample A at $V_g$ = 0 V (electrons, solid squares) and -70 V (holes, open squares). (b) Non-local MR as a function of gate voltage and dc bias current for sample A. (c) Non-local MR as a function of dc bias current for sample B at $V_g$ = 10 V (electrons, solid squares) and -60 V (holes, open squares). (d) Non-local MR as a function of gate voltage and dc bias current for sample B.

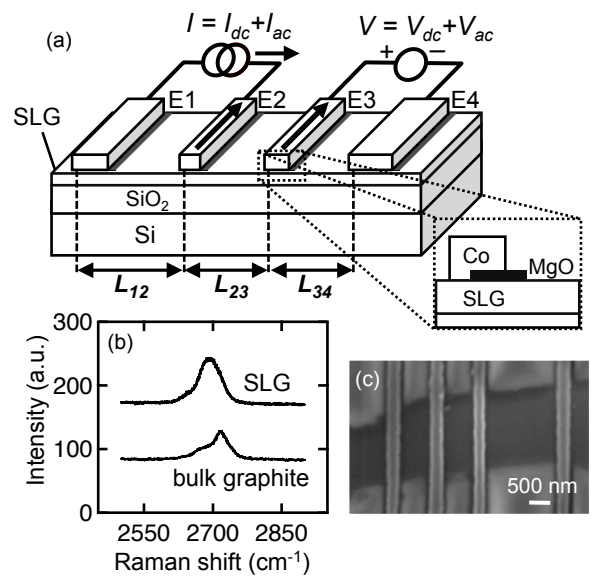

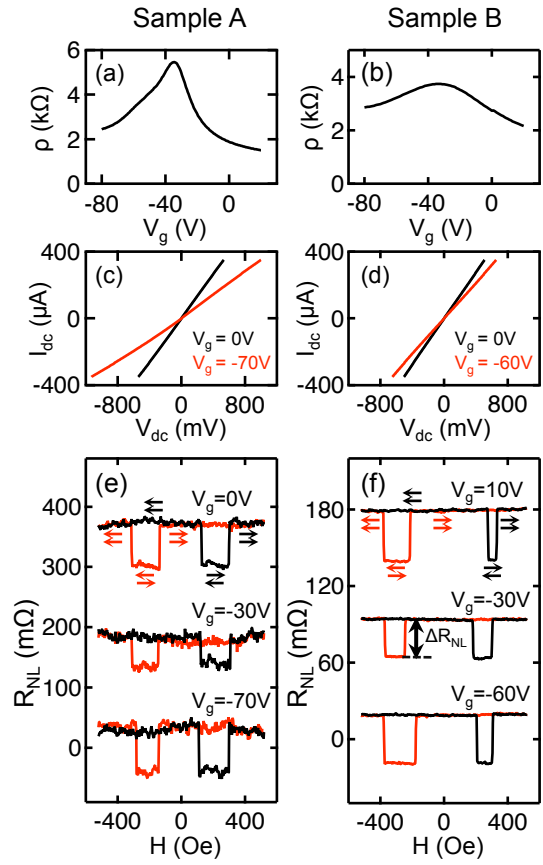

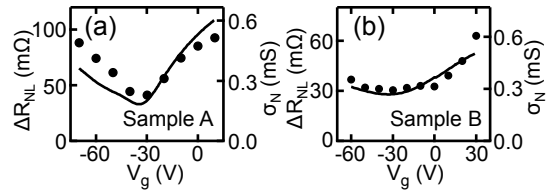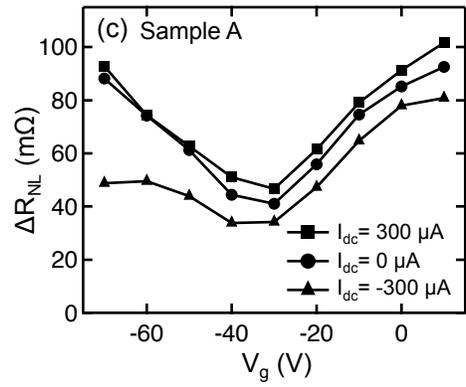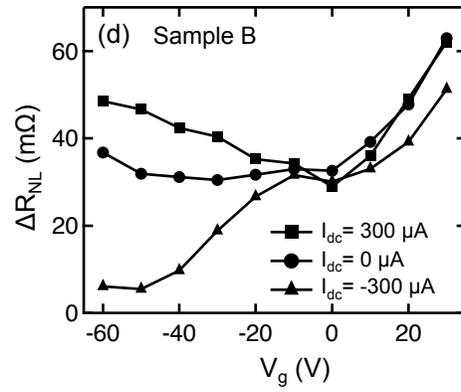

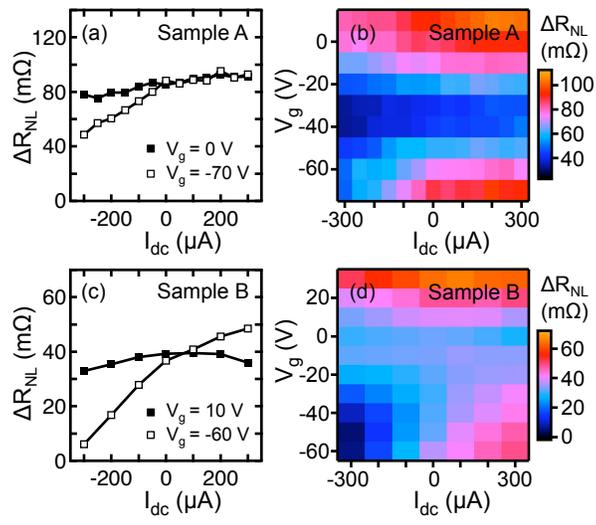